# SECODA: Segmentation- and Combination-Based Detection of Anomalies

Ralph Foorthuis
Data Services / CIO Office
UWV, Amsterdam, the Netherlands
ralph.foorthuis@uwv.nl

*Abstract* — This study introduces SECODA, a novel general-purpose unsupervised non-parametric anomaly detection algorithm for datasets containing continuous and categorical attributes. The method is guaranteed to identify cases with unique or sparse combinations of attribute values. Continuous attributes are discretized repeatedly in order to correctly determine the frequency of such value combinations. The concept of constellations, exponentially increasing weights and discretization cut points, as well as a pruning heuristic are used to detect anomalies with an optimal number of iterations. Moreover, the algorithm has a low memory imprint and its runtime performance scales linearly with the size of the dataset. An evaluation with simulated and real-life datasets shows that this algorithm is able to identify many different types of anomalies, including complex multi-dimensional instances. An evaluation in terms of a data quality use case with a real dataset demonstrates that SECODA can bring relevant and practical value to real-world settings.

## I. INTRODUCTION

*Anomaly detection* (AD) aims at identifying cases that are in some way awkward and do not appear to be part of the general pattern(s) present in the dataset [1, 2, 3, 4, 5, 6]. Such an analysis is often also referred to as *novelty detection* or *outlier identification* [7]. Anomaly detection can be used for various goals, such as fraud detection, data quality analysis, security scanning, process monitoring and data cleansing prior to statistical modelling.

Depending on the specific situation and goals, an anomaly can be taken to mean different things. As [8] put it, "one person's noise is another person's signal". Nonetheless, several types of cases can generally be acknowledged as anomalies. The Theory section presents an overview of these types.

This article presents a novel unsupervised non-parametric anomaly detection algorithm for datasets containing continuous (numerical) and/or categorical attributes. The algorithm in case is SECODA, SEgmentation- and COmbination-based Detection of Anomalies, the core purpose of which is the identification of different types of anomalies. The algorithm is deliberately kept simple for several reasons. First, it is relevant for academia to know that sophisticated AD analysis results can be obtained by relatively simple (and hitherto unexplored) principles and implementations. Second, it makes it possible for practitioners to implement the algorithm on basic platforms, such as machines with relatively little memory and simple DBMS systems that do not offer support for advanced analytics. It also allows for in-database analytics, i.e. analyzing the data in the database itself. This avoids the need to export the data to a separate analytics application, which positively affects time performance and security. The algorithm will therefore be restricted to basic data operations (sort, count, join), control flows (loops) and set-based actions (no point-to-point distances or associations, no complex fitting procedures), so as to show that this can yield sophisticated anomaly detection results.

As part of a real-world evaluation, we will demonstrate how SECODA, and indeed anomaly detection in general, can contribute to improving data quality. In statistics, data quality is evidently important for the analysis process [2, 3]. However, high-quality data is also relevant in broader organizational settings, as it is important for obtaining various kinds of benefits, such as increasing the value of IT systems, enhancing customer service performance, optimizing decision making and improving organizational efficiency [9, 10, 11, 12].

This paper proceeds as follows. The Theory section presents a typology of anomalies, related research and a description of the SECODA algorithm. The Algorithm Evaluation section presents the research approach, results and discussion. The Conclusion summarizes the contributions and discusses further research.

## II. THEORY

### A. Typology of Anomalies

The literature mentions several ways to distinguish between types of anomalies. For example, in sequence or time series analysis, so-called additive, innovational, level shift, and transitory change outliers are often acknowledged [13]. A distinction between weak outliers (noise) and strong outliers (true anomaly) can also be made [1]. In the context of regression analysis it is common to distinguish between outliers, high-leverage points and influential points [2, 3]. More in general, one can differentiate between point, contextual and collective anomalies [14, cf. 15].

The types of anomalies mentioned above are either too specific or too general for the purpose of this study. An alternative typology is therefore presented below. This typology presents a detailed and tangible definition of generic anomaly types. The types are illustrated in figures 1 to 4 (note: the reader might want to zoom in on a digital screen to see colors, patterns and data points in detail).

- *I. Extreme value anomaly*: A case with extremely high or low values on one or multiple individual numerical attributes [cf. 1, 3]. Such a case has one or more values that can be considered extreme when the entire dataset is taken into account. Traditional univariate statistics typically considers this type of outlier, e.g. by using a measure of central tendency plus or minus 3 times the standard or median absolute deviation [16, 3]. The cases with label *Ia* in Fig. 1 are examples, as well as case *Ib* in Fig. 2.
- *II. Sparse class anomaly*: A case with a rare categorical value or a rare combination of categorical values [cf. 17]. This value (or combination thereof) is rare in the entire dataset. Case *IIa* in Fig. 4 is an example, as it is the only green data point in the set.
- *III. Multidimensional numerical anomaly*: A case without extreme values for any of its individual numerical attributes, but which does not conform to the general pattern when multiple numerical attributes are taken into account [cf. 14, 15]. Such cases hide in multidimensionality [2], so several attributes have to be analyzed jointly to detect that they are located in an isolated area. Case *IIIa* in Fig. 1 is an example, as is case *IIIb* in Fig. 3. Data point *IIIc* in Fig. 2 is an illustration as well, especially if the color attribute is ignored.
- *IV. Multidimensional mixed data anomaly*: A case with a categorical value or a combination of categorical values that in itself is not rare in the dataset as a whole, but is only rare in its neighborhood (numerical area). As with type III anomalies, such cases hide in multidimensionality and multiple attributes need thus to be jointly taken into account to identify them. In fact, multiple datatypes need to be used, as a type IV anomaly per definition needs both numerical and categorical data. Cases *IVa* in Fig. 2 and *IVb* in Fig. 4 are anomalies that have a color rarely seen in their respective neighborhoods. Cases can also take the form of second- or higher-order anomalies, with categorical values that are not rare (not even in their neighborhood), but are rare *in their combination* in that specific area. Case *IVc* in Fig. 4 is an example, with both its big size and its red color being normal in that neighborhood. However, its combination of big size and red color renders it anomalous there.

The value of this typology lies not only in providing both a theoretical and tangible understanding of the types of anomalies, but also in evaluating which type of anomalies can be detected by a given algorithm. Interestingly, most research publications do not make it very clear which type of anomaly can be detected [e.g. 17, 18, 19, 20, 21, 22]. For a long time, research has focused mainly on studying the performance of technical aspects such as speed, dataset size and number of attributes, and seems to have largely neglected the functional aspects of AD. However, it is a good practice to provide tangible insight into to the functional capabilities of an algorithm. This paper will therefore also evaluate SECODA by testing which types of anomalies it is able to detect.

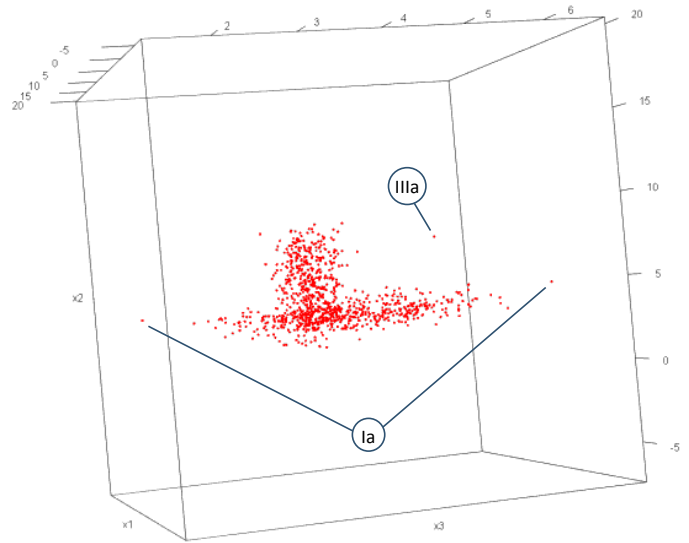

Fig. 1. Dataset "Mountain" with 3 numerical attributes (and 3 labeled anomalies)

*B. Related Research*

A substantive amount of research on anomaly detection has been published [1, 7, 14, 23]. AD originally involved *statistical parametric methods* that focus on univariate outliers [3, 16]. These methods are limited in their practical use since they can analyze only a single numerical attribute that is assumed to conform to a given statistical distribution. Non-parametric multidimensional *distance-based methods*, which can be traced back to [8, 24], were consequently developed. This approach for AD focuses on the distance between individual data points and their nearest neighbors, and has been advanced throughout the years in order to also take into account larger datasets as well as categorical attributes [18, 19, 20, 21, 25]. The anomalies in these methods are the data points that lie furthest from the other cases, using e.g. the Euclidean or Hamming distance. *Density-based approaches*, focusing on the amount of data points in each point's neighborhood, offer a related non-parametric method for detecting outliers [4, 5, 26]. Other modern AD approaches utilize *complex non-parametric statistical models* to identify anomalous cases. Notable examples are One-Class Support Vector Machines [27], ensembles [28, 29] and various subspace methods [1, 30, 49].

The SECODA algorithm presented in this paper is mostly akin to the density-based approach, in which anomalies are the cases located in low-density areas. SECODA employs the so-called histogram-based technique, which is one of the traditional density-based methods for outlier analysis [1, 14]. The core of this technique, often applied in intrusion detection, is determining the frequency of the different types of occurrences. For continuous attributes this implies their discretization into separate bins (i.e. intervals), of which the frequency can be readily determined. Low-frequency cases are considered anomalies. Examples of this technique are the intrusion detection systems described in [31, 32]. Another example is the AVF method, which uses the class frequencies of categorical attributes to detect outliers [17].

Other publications in which frequency-based anomaly detection plays a role are [6, 22, 33, 34, 50, 51]. Histogram-based methods are relatively simple to implement [14] and can generally be expected to have good time performance. A challenge with these methods is determining the optimal arity (i.e. number of discretization intervals) to handle the continuous variables, since the frequency distribution needs to be modeled at a level of granularity that is sufficient to detect true anomalies [1, 14]. Furthermore, these methods run the risk of being too locally oriented and thus neglecting the global characteristics of the dataset, and are prone to the curse of dimensionality [ibid.]. Furthermore, existing methods do not take into account interactions between attributes [14]. These challenges will be addressed in the next section. Additional similarities of SECODA with ensembles, distance- and density-based approaches will be presented in the Discussion section.

*C. The SECODA algorithm*

SECODA advances the traditional histogram-based approach for outlier detection. Discretization of continuous attributes is used in order to be able to jointly take into account both categorical and continuous variables. An *iteratively increased* number of (ever narrower) *discretization intervals* is used to avoid arbitrary and suboptimal bin sizes. It also allows for a detailed and fine-grained analysis of attribute value combinations. We refer to these combinations as *constellations*, which can technically be regarded as concatenations of attribute values. This so-called 'concatenation trick' enables the analysis of datasets containing mixed data, i.e. sets featuring both numerical and categorical attributes. It also ensures that all unique combinations of values can be identified and, if only one or a few cases are instances of such a constellation, are reported as anomalies. Any interaction between attributes will also be captured by the constellations.

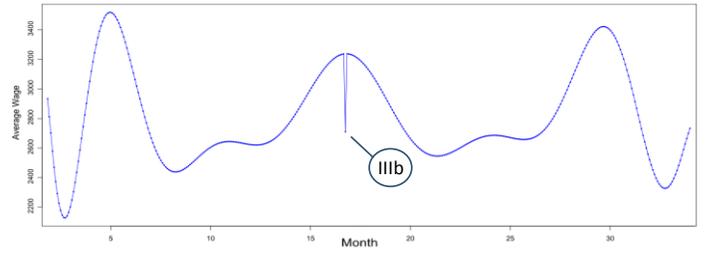

Fig. 3. Dataset "TimeSeries" with 2 numerical attributes, with the points representing a time sequence

Each of the individual attribute values in the next example occur frequently. However, the two constellations "MALE LION / AGE BETWEEN 48 AND 72 MONTHS / MANE" and "MALE LION / AGE BETWEEN 0 AND 12 MONTHS / MANELESS" occur relatively often, whereas the constellation "FEMALE LION / AGE BETWEEN 48 AND 72 MONTHS / MANE" rarely occurs. Because of such inter-variable relationships, it is crucial not to analyze each attribute separately, but to take the interactions between variables into account. The key is to identify the constellations (i.e. the existing combinations of values) and to analyze how frequently they occur.

The SECODA algorithm is presented in Figures 5 and 6. Let $X_i$ represent a $n \times p$ matrix in the $i$th iteration, with $i=0$ being the original full matrix with $n$ rows, and $i>0$ being subsets (in terms of rows) of this original matrix in subsequent iterations. Let $x_{g,h,i}$ represent the matrix value of the $h$th attribute (variable) for the $g$th row (case) in the $i$th iteration, with $g=1,2,\ldots n$ and $h=1,2,\ldots p$. Let $y_i$ represent a column vector with $n$ rows in the $i$th iteration. Let $y_{g,i}$ represent the vector value of the $g$th row (case) in the $i$th iteration, with case $g$ of $y_{g,i}$ referring to case $g$ of $x_{g,h,i}$. The $\oplus$ symbol represents concatenation of multiple attributes into one combined variable. Furthermore, a semi-colon represents the end of a statement, whereas #**green remarks** provide explaining comments. The function noc() returns the number of rows (cases, elements) in a matrix or vector. See the referenced R code for the complete implementation and some additional details.

The process starts by discretizing the continuous attributes into $b=2$ equiwidth bins (i.e. equal interval ranges [6, 54]) in the first iteration. For each case the constellation of which it is an instance is then determined by concatenating all categorical and discretized numerical values. By subsequently calculating the constellation frequencies it can be determined how rare each case is in the current iteration. In the next round this process is repeated with a higher number for $b$. Each case $g$'s average anomaly score $aas_{g,i}$ can be calculated in each iteration $i$ by calculating the arithmetic mean of the case's current constellation frequency and its average score of the previous iteration.

Extreme anomalies will be identified in early iterations. As the process continues and the number of iterations and segmented bins increases, less extreme anomalies will also be isolated in low-frequency constellations and thus get assigned a relatively low average anomaly score. This continuing process will render the cases' scores more precise as well. How does the algorithm know it has *converged* and can stop the process? Before explaining this, it is important to understand that the anomaly score represents the average number of cases similar to case $g$ and is thus a measure of (non-)normalness. Lower scores represent increasingly rare and anomalous cases. To decide whether sufficient anomalies have been identified, the algorithm verifies if a given fraction of the cases in the original dataset has a score below the anomaly threshold. For this study the fraction was set on 0.003. This is based on the often used definition of outliers being those extreme points that deviate at least 3 times the standard deviation from a central tendency measure [6, 16]. In the first iterations the algorithm determines whether there are sufficient cases that can be regarded as truly unique, i.e. sufficient cases being instances of constellations that are observed only once. As the process continues this condition is relaxed, with the stop point (threshold) $s$ starting with 1 and increasing with 0.1 in the first 10 iterations, and increasing with 1 in later iterations. The rationale for this is that, apparently, the situation requires a broader definition of anomalies because it is relatively difficult to find truly unique cases in the given dataset.

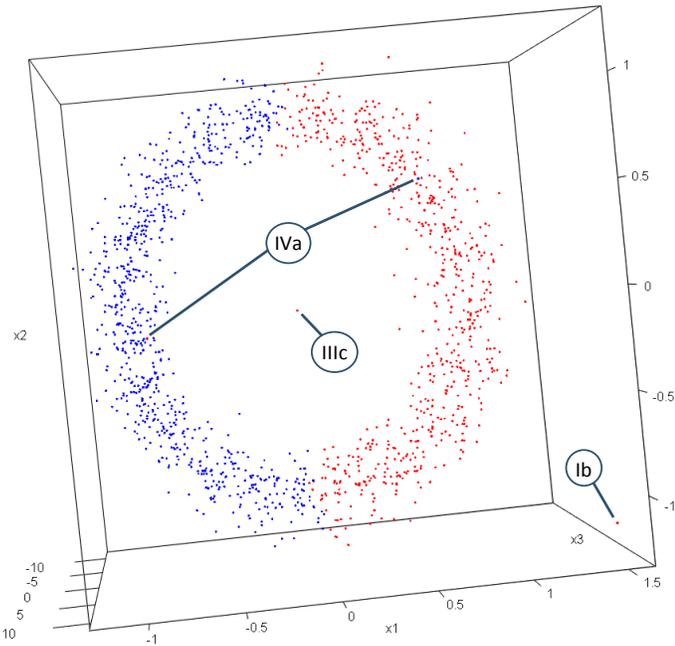

Fig. 2. Dataset "Helix" with 1 categorical (color) and 3 numerical attributes

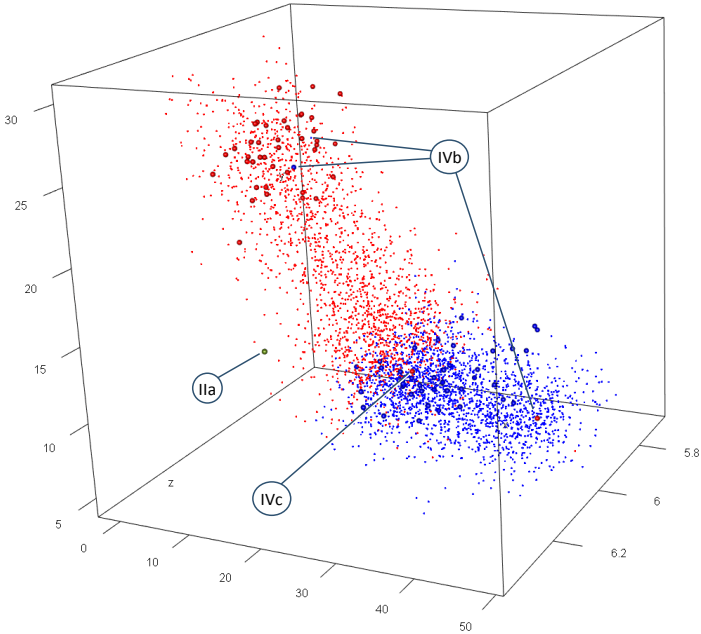

Fig. 4. Dataset "NoisyMix" with 3 numerical attributes and 2 categorical attributes (color and size)

To speed up the analysis, a heuristic is used to *prune* away that part from the search space that, from an anomaly detection perspective, is the least promising and can thus be neglected in any future iterations [cf. 20, 35]. SECODA intrinsically keeps information about the most and least promising cases: the data points with the highest average anomaly scores of a given iteration are the most normal and, at some point, can thus be discarded for the remainder of the AD process. To be safe, this heuristic starts filtering after 10 iterations. The pruning takes the form of retaining the cases with an average anomaly score below the 0.95th quantile (aka the 95th percentile) of the current (potentially already pruned) dataset $D_i$, which in principle means discarding 5% of the cases. While this percentage may appear arbitrary, in practice this constitutes a dynamic and self-regulating pruning mechanism (see the Discussion).

Two other tactics are used to speed up the analysis process. First, the average anomaly score is actually a *weighted* mean, with the frequency of the current iteration weighing the same as the scores of all previous iterations combined. This essentially increases the weights exponentially as the process continues to iterate, resulting in a significant speed-up. The reason for this is the fact that, as the number of bins $b$ increases in later iterations, the number of unique constellations also increases. Therefore, constellation frequencies generally drop and thus drag the individual average anomaly scores down – especially for anomalies, as these imply low-frequency constellations. This score thus drops faster if later iterations have an exponentially higher weight, resulting in a quicker convergence. The weights also prevent bias, which will be explained in the Discussion section. As a second tactic for speeding up the analysis, after 10 iterations $b$ is increased each time with *larger steps*. As cases get increasingly isolated from one another, the number of unique constellations increases faster. This results in an accelerated version of the mechanism described above, and thus yields an even faster convergence.

### III. ALGORITHM EVALUATION

Anomalies, by definition, are rare and labeled anomalies are usually not available. Therefore, both qualitative and quantitative methods for evaluating anomaly detection algorithms are used in research [1]. This study presents three types of evaluations of SECODA. First, simulated (synthetic) datasets are used to study whether the algorithm is capable of identifying the different types of anomalies presented in section II. Second, we use two real-world datasets with labeled anomalies to evaluate SECODA with ROC/PRC curves and related performance metrics. Third, the results of a real-world data quality use case are presented.

#### A. Experimental setup and Datasets

Table I provides an overview of the simulated and real-world datasets that were used to evaluate the algorithm.

TABLE I. DATASETS USED FOR EVALUATION

| Dataset | Nature | Datatypes | # Cases | Types of anomaly |
|---|---|---|---|---|
| Mountain (Fig. 1) | Simulated | 3 numerical | 943 | Type I, Type III |
| Helix (Fig. 2) | Simulated | 3 num, 1 categ | 1410 | Type I, III, IV |
| TimeSeries (Fig. 3) | Simulated | 2 numerical | 398 | Type I, Type III |
| NoisyMix (Fig. 4) | Simulated | 3 num, 2 categ | 3867 | Type II, Type IV |
| Polis dataset 1 (Fig. 9) | Real-world data | 3 num, 1 categ | 162980 | Type I, II, III, IV |
| Polis dataset 2 | Real-world data | 3 num, 1 categ | 1063076 | Type I, II, III, IV |

The simulated datasets were generated in R and are depicted in Fig. 1 to 4. The real-world data were drawn randomly from the Polis Administration, which constitutes an official register maintaining masterdata regarding the salaries, social security benefits, pensions and income relationships of people working or living in the Netherlands. This register is owned by three institutions collaborating in an award-winning strategic alliance, namely the Dutch Tax and Customs Administration, Statistics Netherlands and UWV, and is maintained by the latter [12, 36]. Using Polis' PLM engine (a parameterized data extraction module), two representative samples were drawn and subsequently anonymized. A data point or case in these datasets represents an income relationship, of which individual citizens can have one or more (e.g. because they receive both a salary and social benefits). Both rule-based verifications and unsupervised anomaly detection are relevant for maintaining the Polis Administration, e.g. for data quality analysis and fraud discovery [12, 37].

The experiments were run on a Windows machine, R 3.3.2, RStudio 1.0.136 (packages pROC 1.9.1, precrec 0.6.2, boot 1.3-18, rgl 0.98.1 and e1071 1.6-8) and self-authored R-code.

**Algorithm: SECODA**
**Inputs:** $D_0$, the original matrix with $n$ cases and $p$ attributes.
**Output:** $aas_i$, a vector of average anomaly scores after the last iteration for all cases in $D_0$, with $aas_{g,i}$ representing the individual score.
**Key local vars:** $b$, the number of discretization bins (arity).
$s$, used as stop point and for increased binning.
$cf_{g,i}$, the current frequency in iteration $i$ of the constellation to which case $g$ belongs.
**begin**
   $i \leftarrow 0; b \leftarrow 2; s \leftarrow 1; continue \leftarrow$ TRUE  # Set initial values
   **while** $continue =$ TRUE **do**
      $i \leftarrow i + 1$
      $D' \leftarrow D_i$ with numerical attributes discretized into $b$ equiwidth bins
      $cf_{g,i} \leftarrow$ ConstellationFrequencyPerCase($D'$)
      **if** $i > 1$  # Calculate average anomaly scores for cases in $D_i$
         $aas_{g,i} \leftarrow \frac{1}{2}(aas_{g,i-1} + cf_{g,i})$
      **else**  # If it's the first iteration, put in the frequency
         $aas_{g,i} \leftarrow cf_{g,i}$
      **end if**

      **if** $i \leq 10$  # Iteration management
         $s \leftarrow s + 0.1$
         $b \leftarrow b + 1$
      **else**  # Take larger steps and prune cases in higher iterations
         $s \leftarrow s + 1$
         $b \leftarrow b + (s - 2)$
         # Add to $aasp_i$ the anomaly scores of the 5% most normal cases that are to be pruned away:
         $p \leftarrow$ subset of $aas_i$, with each $aas_{g,i} \geq 0.95$ quantile value
         $aasp_i \leftarrow aasp_{i-1} \cup p$
         # Prune away high-frequency (normal) cases for next iteration:
         $D_{i+1} \leftarrow$ subset of $D_i$, with each case such that its $aas_{g,i} < 0.95$ quantile value
      **end if**

      $Q \leftarrow$ Subset of $D_i$, with each case such that its $aas_{g,i} \leq s$
      **if** (noc($Q$) / noc($D_0$)) > 0.003  # Verify fraction of identified anomalies
         $continue \leftarrow$ FALSE  # No new iteration (process has converged)
      **end if**

   **end while**
   $aas_i \leftarrow aas_i \cup aasp_{i-1}$  # Combine average anomaly scores from latest iteration with scores from cases that have been pruned previously
   **return** $aas_i$  # Return full anomaly score vector as the end result
**end**

Fig. 5. The SECODA algorithm

### B. Identification of the different types of anomalies

Four simulated datasets were used to study whether SECODA was able to identify the different types of anomalies presented in section II. The four sets are presented in Table I and visually depicted in Figures 1 to 4. SECODA was able to identify all types of anomalies. Anomaly *IIIa* in Figure 1 was found to be the number one anomaly in the Mountain dataset, while the two *Ia* anomalies were ranked third and fourth. Figure 2's two *IVa* cases were ranked as the first and second anomalies of the Helix dataset, *IIIc* was ranked third, and anomaly *Ib* was reported as the fourth anomaly. If only the three numerical variables are provided as input to SECODA, case *Ib* was reported as the first and case *IIIc* as the second anomaly (the *IVa* cases could obviously not be identified because Type IV anomalies only exist in sets containing mixed data). Interestingly, although using multiple kernels, One-Class Support Vector Machines were not able to detect the *IIIc* anomaly in the Helix dataset, regardless of whether or not the categorical attribute was provided as input (as a dummy variable). SECODA reported Figure 3's *IIIb* data point as the number one anomaly. Figure 4's *IIa* case and the two upper *IVb* cases shared the number one rank as anomalies. This is the result of these cases being unique data points in all iterations of the detection process, resulting in an average anomaly score of exactly 1. Case *IVc* and the lower *IVb* case were reported as the fourth and fifth anomaly. If only the two categorical attributes are provided as input for SECODA, the *IIa* anomaly is reported as the number one (and only) anomaly. This is as expected, since this case is a Type II *sparse class* anomaly in a set with only categorical data.

### C. Real-life dataset with ground-truth

This section evaluates different versions of the SECODA algorithm on the Polis datasets. Fig. 7 presents the time performance of 3 SECODA versions. For each version 5 random subsets of Polis dataset 2 were created (from 1/5$^{th}$ to 5/5$^{th}$ of the 1,063,076 cases). Each of these 15 combinations was analyzed by taking the arithmetic mean of 100 samples, resulting in a total of 1500 SECODA runs. The red line represents the final algorithm, blue represents SECODA without pruning ("Pruneless"), and green represents SECODA without increasing the number of intervals $b$ with larger steps ("Stepless"). The index of 100 is based on the fastest average run, which is 6.9832 seconds. As can be seen, the final version indeed performs fastest and scales linearly with dataset size. The pruning results in the largest speed-up, although the increased stepping is almost as effective. Together the two tactics result in a combined time performance improvement.

These results beg the question of whether the final algorithm, with its effective heuristics to speed up the analysis, performs as well as the other versions with regard to identifying true anomalies. The results of a test with Polis dataset 1 are presented below. Existing data quality rules – developed totally independently of this study and not by this author – were used to create a test set with labeled data, yielding 106 anomalies on the set of 162,980 cases. These rules do not necessarily identify cases that are guaranteed to be incorrect, but all of the positive cases can be regarded as true anomalies in the context of this study.

Since SECODA returns a full set of gradual anomaly scores rather than binary TRUE/FALSE labels, ROC and PRC curves and their respective AUCs are the appropriate methods to compare the functional performance of the various versions of the algorithm [1, 38, 39].

**Algorithm:** ConstellationFrequencyPerCase
**Inputs:** $D'$, containing $p$ (categorical and discretized numerical) attributes and a total of $n$ cases, with $n \leq noc(D_0)$.
**Output:** $cf_i$, a vector with for each case $cf_{g,i}$ the frequency of the constellation to which the case belongs in the current iteration.
**begin**
    # Concatenate each case's attribute values in this iteration (i.e. determine the constellations):
    $cc_{g,i} \leftarrow d'_{g,1,i} \oplus d'_{g,2,i} \oplus ... \oplus d'_{g,p,i}$
    # Determine the frequency of distinct constellations in this iteration (with $k$ identifying the constellations):
    $ccf_{k,i} \leftarrow$ The number of cases per constellation
    # Determine the frequency of each case, using the frequencies of their constellations (i.e. inner join $cc_i$ and $ccf_i$ on $k$):
    $cf_{g,i} \leftarrow$ The frequency from $ccf_{k,i}$ for each case's corresponding constellation
    **return** $cf_i$  # Return each case's current frequency $cf_{g,i}$ as the elements of a vector
**end**

Fig. 6. The SECODA algorithm

The 95% confidence intervals (CIs) of these metrics are calculated with 10000 stratified percentile bootstrap resamples, using the vertical averaging method for the CIs of the ROC curve in Fig. 8 [38, 40].

Table II and Fig. 8 present the ROC curves and metrics of different SECODA versions. As can be seen from the figure, the ROC curves overlap. Moreover, the inset shows that the 95% confidence intervals (vertical averaging, shown in orange) of the final SECODA version includes the two other ROC curves. The confidence intervals of the other SECODA versions and of the specificity dimension show a very similar pattern. The so-called confidence bands for entire curves can be expected to be even wider [41].

Table II presents the AUCs of the ROC and PRC of the 3 evaluated SECODA algorithms. Since there are few anomalies and many normal (i.e. negative) instances, the performance in the far left-hand side of the ROC graph is interesting to study in more detail [38]. Therefore, a standardized partial AUC with specificity restricted between 100% and 90% is regularly used [40, 42, 43, 52] and also included here. As can be seen, in all cases the confidence intervals of a given AUC include the AUCs of the other versions (this is also true for DeLong confidence intervals, which can be used only for full AUCs).

Three hypothesis tests were performed to also compare the three 100-90% partial ROC AUCs in a pair-wise manner. None of the bootstrapped P-values of the combinations in Table II are statistically significant, indicating that they should be considered to be performing equally well.

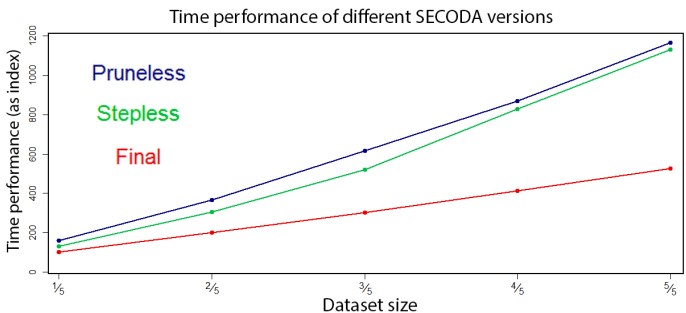

Fig. 7. Time performance of the different SECODA versions on Polis set 2

This all consistently leads to the conclusion that no significant differences exist between the individual curves, nor between the AUCs. It can therefore be concluded that the different heuristics used to boost the time performance of SECODA do not have an adverse effect on its functional ability to detect true anomalies.

Table III presents the final algorithm's metrics for the optimal threshold according to the Youden index and Matthews Correlation Coefficient [46, 53]. Due to the imbalanced distribution (i.e., few anomalies and many normal cases), several metrics are naturally high. Conversely, the precision is low because the Youden threshold does not take this metric into account, while the Kappa is low because it is adjusted for chance. The F1 measure also is relatively low, as this is a harmonic mean and will thus not have a high value if either precision or sensitivity (recall) is low. In any case, the table shows that large differences exist between thresholds and that it is thus important to conduct a ROC/PRC analysis. A practitioner should obviously base his or her decision regarding the optimal threshold on the requirements of the specific situation. In an exploratory data analysis an explicit threshold may not even be necessary. The gradual score allows simply scrutinizing the top $x$ anomalies, depending on time and budget constraints.

*D. Real-life dataset - qualitative inspection of data quality*

This section presents a manual inspection of the AD results found by SECODA. For this, the 30 most extreme anomalies were scrutinized. Upon visual inspection and by ignoring any domain knowledge, these can indeed all be considered anomalies. All anomaly types of section II were detected. Fig. 9 presents the 30 anomalies identified by SECODA as large dots, and includes examples of each type. The Polis dataset 1 includes three continuous attributes that represent sums of money (e.g. income and sums withheld for social security) as well as a variable that represents a social security code. Ignoring the 30 anomalies for a moment, the 4D plot of raw data already is indicative of high-quality data. Such strict patterns, caused i.a. by laws and regulations, would not be visible otherwise. However, anomalies can nonetheless be found. The 2 blue *Ic* data points represent *extreme value anomalies*. Point *IIb* is one of very few orange points and thus a *sparse class anomaly*. The *IIId* cases can be regarded as *multidimensional numerical anomalies*, as these points lie somewhat isolated from the pattern but have no extreme values. Finally, cases *IVd* are clear instances of *multidimensional mixed data anomalies*, as they are blue points in an otherwise pink local pattern (or vice versa).

Taking domain knowledge into account, several of the anomalies identified by SECODA proved to be indicative of previously unknown data quality problems. These anomalies referred to records that had a missing value for the categorical attribute (which did not occur very often in some areas and were thus found to be anomalous there). These missing values presented themselves as an additional code that formed input for the AD analysis. Upon further inspection and discussion with Polis experts it was found that it was both possible and desirable to avoid these missing codes when data is exported from the database. The empty cells in the export were the result of data selection from different income-related entities (database tables), each of which features its own time dimension (to serve different stakeholders and to ensure storage efficiency). In certain cases the entities' starting dates differed, yielding several empty cells in the combined data file. Although the existing data storage and PLM export were both correct from a technical database perspective, the extracted data file was not optimal from a semantic data (quality) perspective. As a result of this AD analysis it was therefore decided to modify the PLM engine, so as to enable a more sophisticated selection and thus to improve the quality of future data deliveries.

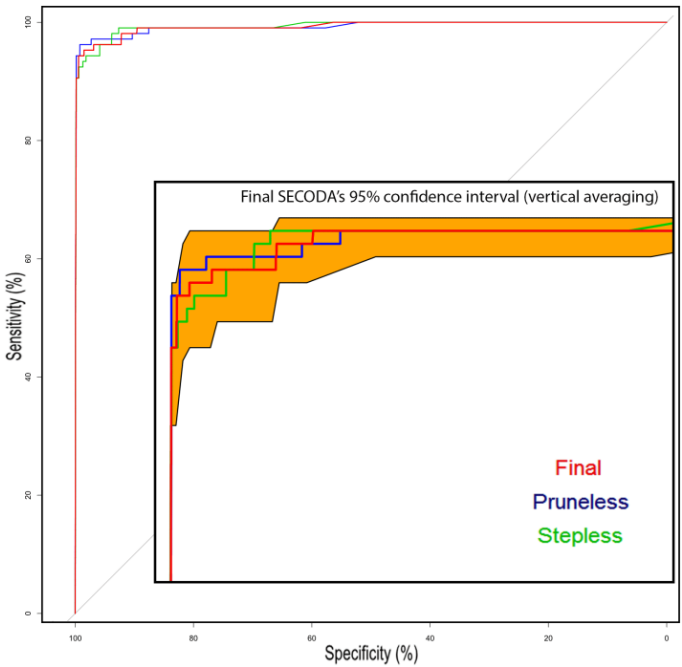

Fig. 8. ROC curves for three SECODA versions (inset: CI for final version)

### E. Discussion

This section discusses various characteristics of SECODA. As opposed to many other AD approaches [e.g. 4, 5, 8, 21, 25], there is no need for SECODA to calculate point-to-point distances or associations. The core of the algorithm identifies constellations by concatenating attribute values and determines the constellation frequencies to investigate how rare individual cases are. The distance concept is only implicitly present in SECODA, namely by iteratively narrowing the discretization intervals (bins). This lets the algorithm analyze increasingly smaller and more detailed segments of the dataset's attributes. Starting with a small number of bins (and thus a global focus) results in anomalies with relatively large distances from similar cases being detected in an early iteration, while more subtle (local) anomalies will be discovered later and will therefore get assigned higher average anomaly scores. This is due to the fact that extreme anomalies will, in early iterations, already belong to constellations with a low frequency. This frequency will remain low in later iterations, resulting in these cases having a low final average score. The core part of the method basically consists of counting, without any need for point-to-point calculations (the basic solution of which results in exponential complexity). The algorithm thus has *low memory requirements*.

SECODA can deal with *complex interactions* between variables, resulting in the ability to also identify Type III and first- and higher-order Type IV anomalies. This is the result of SECODA's focus on constellations. The cases with a rare combination of attributes will belong to a constellation featuring a low frequency. This will be the case in multiple iterations and the overall average anomaly score will thus be low.

The *concatenation trick* affords other beneficial properties as well. Because the constellations technically are concatenations of different data types into a single string, SECODA can easily process mixed data. This also allows the algorithm to automatically deal with missing values, as these can be represented in a concatenation without problems. Moreover, missing values are automatically handled as one would functionally desire in an AD context, with very sparse missing values yielding low average anomaly scores (representing anomalies) and high-frequent missing values resulting in relatively high scores. This was observed in the Polis analysis. In the same vein, multicollinearity does not pose a problem and there is no need to standardize the numerical attributes before running the algorithm. Working with constellations does nothing, however, to alleviate the curse of dimensionality.

Some other analytical properties of SECODA are worth discussing as well. It has already been mentioned that, to *speed up* the analysis, the anomaly score is an average frequency score. It should be stressed that this is a *weighted* average, with the weights increasing *exponentially*. As a result, later iterations (and thus the local neighborhood of the data point) have a relatively strong influence (although global influences are retained). Alternatively, the score can be interpreted as a non-parametric density measure with respect to the entire dataset, but focusing especially on the immediate neighborhood due to the weighted average.

TABLE II. PERFORMANCE METRICS OF DIFFERENT SECODA VERSIONS

|  | SECODA version | | |
|---|---|---|---|
|  | **Final** | **Pruneless** | **Stepless** |
| **ROC AUC (95% CI)** | 99.2472101% (98.3030214%-99.8237980%) | 99.2612359% (98.2368757%-99.8936765%) | 99.2934346% (98.4734947%-99.7952676%) |
| **ROC partial AUC for 100-90% specificity (95% CI)** | 97.5908842% (95.6997905%-99.0829036%) | 97.9739000 (96.1351669%-99.4404026%) | 97.5786719% (95.9015903%-98.9350227%) |
| **ROC partial AUC for 100-90% sensitivity (95% CI)** | 96.3403181% (91.4859672%-99.3746080%) | 96.4181925% (91.0116888%-99.7348309%) | 96.5967076% (92.5246381%-99.2143688%) |
| **PRC AUC (95% CI)** | 99.9994304% (99.9986572%-99.9998829%) | 99.9994193% (99.9985530%-99.9999290%) | 99.9994811% (99.9988207%-99.9998658%) |
| **P-value (two-sided) of pair-wise partial AUC difference test for 100-90% specificity** | With Pruneless: P = 0.2028067 | With Stepless: P = 0.2960681 | With Final: P = 0.9628931 |

TABLE III. PERFORMANCE METRICS OF FINAL SECODA ALGORITHM

| Metric | Best Matthews CC threshold (5.3622) | Best Youden ROC threshold (42.4671) |
|---|---|---|
| Sensitivity/Recall | 0.9056604 | 0.9528302 |
| Specificity | 0.9984405 | 0.9852401 |
| Precision/PPV | 0.2742857 | 0.0403194 |
| Accuracy | 0.9983802 | 0.9852190 |
| F1 measure | 0.4210526 | 0.0773651 |
| Matthews CC | 0.4979220 | 0.1944046 |
| Cohen's Kappa | 0.4204740 | 0.0762121 |

The exponentially increasing weights also *prevent bias*. An unweighted analysis may suffer from bias when the data distribution is skewed due to one or several extreme value anomalies on one side. As a result of the arbitrary discretization cut point(s), the first iteration(s) may assign a small part of high-density data points to the same low-frequency bin as these Type I anomalies. Later iterations will correct this by creating ever smaller bins, but for unweighted analyses it was found that this was generally not corrected before the algorithm converged. Although Type I and II anomalies were still detected correctly, the identification of Type III and IV anomalies could suffer severely. The weighted version of the algorithm corrects the bias swiftly, however, because the influence of later iterations increases exponentially. Because of its functional bias, the unweighted version of SECODA was not included in the evaluation in section III.C. Finally, it is worth noting that the weighted average not only serves to speed up the analysis and prevent bias, but also ensures that a minimum amount of information has to be passed on to new iterations. On case-level only the latest average anomaly score needs to be retained, not all iterations' frequencies. This aids in keeping a *low memory imprint*.

When analyzing the anomalies of Polis dataset 1, it proved to be very valuable to also inspect *multidimensional plots*, such as displayed in Fig. 9. Studying the attribute values of alleged anomalies by using a data table does not necessarily provide tangible insights into why a case is denoted as such, especially not for multidimensional anomalies. However, a visual depiction may provide the necessary contextual information, as one can see the patterns in the data – and how the anomaly does not fit in. However, note that there could be more dimensions that can be conveniently visualized. Moreover, even visual plots do not always provide direct cues, since they usually do not show the number of cases present on the same numerical location (a heatmap could be employed, but this implies that the color dimension cannot be used anymore for categorical attributes). The number of cases in given areas (density) is very important, however, for the scores that individual cases ultimately get assigned by the algorithm. For example, a high number of cases concentrated in a single point in the data space (e.g. in the origin) may result in a high end score for a case that is relatively nearby but otherwise visually isolated. This is a direct consequence of the fact that its relative density is still quite high. This may not always be readily visible from a plot.

The pruning heuristic is also worth discussing. From a design perspective the algorithm should discard exactly the 5% most normal cases of the current iteration's dataset. However, the experiments show this heuristic to be a *self-regulating mechanism* during runtime. It dynamically decides how many cases to discard, depending on the number of similar cases in that particular iteration. The percentage of cases discarded is often significantly more than 5% as a result of the heuristic's focus on the most normal cases. If many cases around the 0.95 quantile have an identical score (which normal cases indeed have, partly due to the discretization), all the cases with that score will be discarded. For example, in many iterations more than 20% of the cases of Polis dataset 1 were discarded, resulting in effective pruning runs.

The convergence criterion, the fraction of 0.003, is based on the literature [6, 16]. Its main function is to decide whether the algorithm has identified sufficient cases that may be anomalies, which practically means that the algorithm has run sufficiently long to isolate enough cases below the score threshold (which is relaxed as the process continues). This implies that the algorithm has obtained sufficiently precise average anomaly scores at that moment, at least for the rare cases. The exact fraction is thus not critical, as it mainly determines the *precision* for distinguishing between less extreme anomalies.

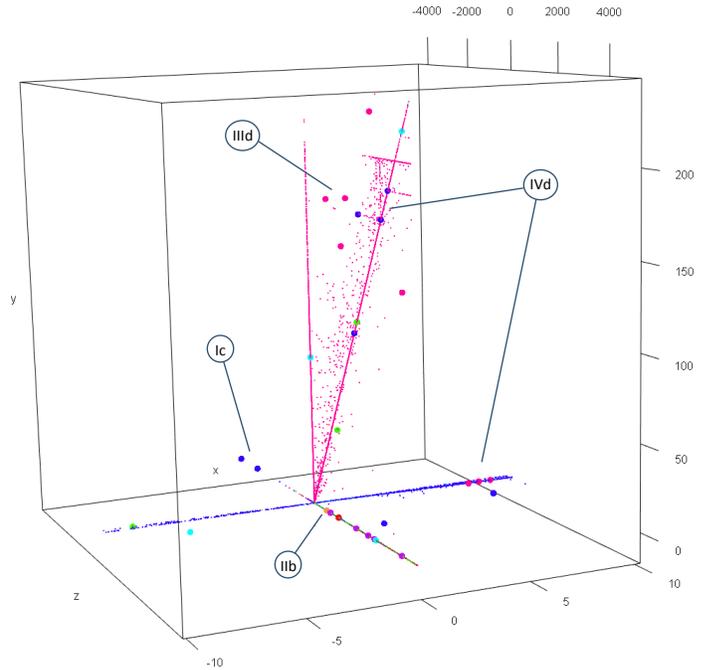

Fig. 9. The top 30 anomalies identified by SECODA for Polis dataset 1

SECODA, using an iterative AD process of which the partial results get averaged, bears *similarities* with ensembles [28, 29]. More specifically, the algorithm can be compared to iForest [29]. Just like this AD method, SECODA identifies anomalies by iteratively searching for cases which can be easily (i.e. with relatively few binning cut points) isolated from the other cases (i.e. of which the constellation has a low frequency). Also, both SECODA and iForest focus on isolating candidate anomalies, and do not attempt to model the 'normal' cases. In this respect SECODA not only features a pruning heuristic, but its stop criterion also depends on a given percentage of rare cases (both mechanisms result in normal cases not being analyzed in detail). However, there are also important differences. SECODA does not use trees and path lengths, and offers the advantage of being able to analyze datasets with mixed data formats.

The SECODA approach can also be likened to the method for high-dimensional outlier detection presented in [6]. Their approach too utilizes discretization, an iterative algorithm and a combinatorial focus. However, there are also fundamental differences, as the solution of [6] iterates over the attributes using one fixed discretization range (while SECODA iterates over the number of bins using one combined view on the attributes). Moreover, they use equidepth (equal frequency) discretization (while the standard SECODA method employs equiwidth discretization), and focus only on numerical variables (while SECODA is able to analyze mixed data).

To conclude this discussion it is valuable to state that the real-world use case with data from the Polis Administration demonstrates that SECODA can contribute to several well-known data quality aspects [cf. 12, 44, 45]. More specifically, this refers to the *correctness* of individual values (the missing values that proved to be incorrect), the *completeness* of cases (a missing value implies incompleteness) and the *consistency* between attribute values (codes in a certain numerical area proved to be suspicious and were detected by SECODA because these cases were scarce in that specific area and thus inconsistent with the other cases).

## IV. CONCLUSION

This article offers multiple contributions. First, a new general-purpose unsupervised non-parametric combination-based anomaly detection algorithm for mixed data is presented. From a theoretical perspective this advances the traditional histogram (density) based approach for AD. The iterative concatenations proved to be a solution for many of the shortcomings known to be present in the histogram-based method, such as a single sub-optimal discretization interval and arbitrary binning, neglecting global characteristics of the dataset, and being unable to detect interactions between attributes. Iteratively (i.e. with multiple discretization intervals) combining the cases' attribute values in a constellation of which the frequency can easily be determined proved to be key here: a single arbitrary interval is avoided, discretization error is reduced, searches are conducted both globally and locally, complex relationships between variables are captured, and rare combinations are readily identified as anomalies – including anomalies hidden in complex multivariate interactions. As a second contribution it is shown that complex anomalies can be detected by a relatively simple algorithm using basic data operations – i.e. based on counting and without point-to-point calculations or complex fitting procedures. This also has practical relevancy, as the simplicity and modest memory requirements should allow for in-database analytics and the analysis of very large datasets. The typology of anomaly types is the third contribution. This typology can be used for theoretical understanding of the nature of different types of anomalies and for the evaluation of AD algorithms. A fourth contribution is the presentation of a real-world case showing that SECODA, and anomaly detection in general, can be used in practice to improve data quality.

There are several opportunities for future research. First, the time performance can be further optimized. The time scaling is linear, but the algorithm can get more competitive for very large datasets when the number of required iterations can be brought down. For example, in situations in which a binary TRUE/FALSE decision is more important than gradual AD scores, the optimal number of binning iterations could be determined before running SECODA. This number could possibly be based on dataset characteristics such as its size, dimensionality, data types and variance. Loss of detection power should be kept to a minimum when further optimizing time performance. A second research opportunity is having the algorithm deal effectively with the curse of dimensionality [cf. 35]. One potential solution may be to start the process by applying dimensionality reduction techniques, such as PCA or multidimensional scaling. However, this brings with it the risk of losing anomalies before even running the AD analysis [46]. Another potential solution could be the random subspace method a.k.a. attribute bagging [47, 48]. This approach, usually applied in ensemble learning, uses random sampling of variables to prevent the many attributes essentially turning every case into an isolated anomaly. A related approach is the evolutionary algorithm of [6], which uses random subsets of the attributes in the dataset to deal with high-dimensionality. To what degree such extensions of SECODA can still guarantee that all unique combinations will be identified should also be studied as part of that research. Dealing with the challenges mentioned above may also imply reverting to more advanced solutions and defy the principle to keep the algorithm implementation restricted to basic data(base) operations. A final suggestion for future research is putting the focus back on the functional aspects of anomaly detection. Since this has been neglected the past years in favor of more technical topics, it can be argued that it is here that most of the value for data quality, fraud detection and other practical applications for AD will lie in the years to come.

**Remarks**. A SECODA implementation for R and several examples are available for download from www.foorthuis.nl. Also see [55].

**Acknowledgments**. The author thanks Safar Niamat, Emma Beauxis-Aussalet, Peter Brom, Bert Voorbraak, Marco Kleinherenbrink for their contributions.

## REFERENCES

[1] C.C. Aggarwal, "Outlier Analysis," New York: Springer, 2013.
[2] A.J. Izenman, "Modern Multivariate Statistical Techniques: Regression, Classification, and Manifold Learning," Springer, 2008.
[3] J. Fielding, N. Gilbert, "Understanding Social Statistics," London: Sage Publications, 2000.
[4] M.M. Breunig, H. Kriegel, R.T. Ng, J. Sander, "LOF: Identifying Density-Based Local Outliers," Proceedings of the ACM SIGMOD Conference on Management of Data, 2000.
[5] S. Papadimitriou, H. Kitagawa, P.B. Gibbons, C. Faloutsos. "LOCI: Fast Outlier Detection Using the Local Correlation Integral," ICDE-03, IEEE 19th International Conference on Data Engineering, 2003.
[6] C. C. Aggarwal, P.S. Yu, "An Effective and Efficient Algorithm for High-Dimensional Outlier Detection," The VLDB Journal, Vol. 14, No. 2, pp 211–221, 2005.
[7] M.A.F. Pimentel, D.A. Clifton, L. Clifton, L. Tarassenko, "A Review of Novelty Detection," Signal Processing, Vol. 99, pp. 215-249, 2014.
[8] E.M. Knorr, R.T. Ng, "Algorithms for Mining Distance-Based Outliers in Large Datasets," VLDB-98, Proceedings of the 24rd International Conference on Very Large Data Bases, 1998.

[9] B.H. Wixom, P.A. Todd, "A Theoretical Integration of User Satisfaction and Technology Acceptance," Information Systems Research, Vol. 16, No. 1, pp. 85–102, 2005.

[10] N. Gorla, T.M. Somers, B. Wong, "Organizational Impact of System Quality, Information Quality, and Service Quality," Journal of Strategic Information Systems, Vol. 19, No. 3, pp. 207-228, 2010.

[11] P. Setia, V. Venkatesh, S. Joglekar, "Leveraging Digital Technologies: How Information Quality Leads to Localized Capabilities and Customer Service Performance," MIS Quarterly, Vol. 37, No. 2, 2013.

[12] P. Daas, S. Ossen, R. Vis-Visschers, J. Arends-Tóth, "Checklist for the Quality Evaluation of Administrative Data Sources," Discussion Paper, CBS, Statistics Netherlands, ISSN 1572-0314, 2009.

[13] R. Kaiser, A. Maravall, "Seasonal Outliers in Time Series," Universidad Carlos III de Madrid, working paper number 99-49, 1999.

[14] V. Chandola, A. Banerjee, V. Kumar, "Anomaly Detection: A Survey," ACM Computing Surveys, Vol. 41, No. 3, 2009.

[15] X. Song, M. Wu, C. Jermaine, S. Ranka, "Conditional Anomaly Detection," IEEE Transactions on Knowledge and Data Engineering, Vol. 19, No. 5, pp. 631-645, 2007.

[16] C. Leys, C. Ley, O. Klein, P. Bernard, L. Licata, "Detecting Outliers: Do Not Use Standard Deviation Around the Mean, Use Absolute Deviation Around the Median," Journal of Experimental Social Psychology, Vol. 49, No. 4, pp. 764-766, 2013.

[17] A. Koufakou, E.G. Ortiz, M. Georgiopoulos, G.C. Anagnostopoulos, K.M. Reynolds, "A Scalable and Efficient Outlier Detection Strategy for Categorical Data". Proceedings of the International Conference on Tools with Artificial Intelligence (ICTAI), 2007.

[18] A. Ghoting, S. Parthasarathy, M.E. Otey, "Fast Mining of Distance-Based Outliers in High-Dimensional Datasets," Proceedings of the 2006 SIAM International Conference on Data Mining, pp. 609-613, 2006.

[19] C.H.C. Teixeira, G.H. Orair, W. Meira Jr., S. Parthasarathy, "An Efficient Algorithm for Outlier Detection in High Dimensional Real Databases," Technical report, University of Minas Gerais, Brazil, 2008.

[20] G.H. Orair, C.H.C. Teixeira, W. Meira Jr., Y. Wang, S. Parthasarathy, "Distance Based Outlier Detection: Consolidation and Renewed Bearing," Proceedings of the VLDB Endowment, Vol. 3, No. 2, 2010.

[21] M. Sugiyama, K.M. Borgwardt, "Rapid Distance-Based Outlier Detection via Sampling," NIPS'13 Proceedings of the 26th International Conference on Neural Information Processing Systems, pp. 467-475, 2013.

[22] C. Krügel, T. Toth, E. Kirda, "Service Specific Anomaly Detection for Network Intrusion Detection," Proceedings of the 2002 ACM Symposium on Applied Computing, pp. 201-208, 2002.

[23] M. Gupta, J. Gao, C.C. Aggarwal, J. Han, "Outlier Detection for Temporal Data: A Survey," IEEE Transactions on Knowledge and Data Engineering, Vol. 25, No. 1, 2014.

[24] E.M. Knorr, R.T. Ng, "A Unified Notion of Outliers: Properties and Computation," KDD-97, Proceedings of the Third International Conference on Knowledge Discovery and Data Mining, 1997.

[25] S.D. Bay, M. Schwabacher, "Mining Distance-Based Outliers in Near Linear Time with Randomization and a Simple Pruning Rule," Proceedings of the Ninth ACM SIGKDD, pp. 29-38, 2003.

[26] J.H.M. Janssens, E. Postma, "One-Class Classification with LOF and LOCI: An Empirical Comparison," Proceedings of the 18th Annual Belgian-Dutch Conference on Machine Learning, pp. 56-64, 2009.

[27] B. Schölkopf, R. Williamson, A. Smola, J. Shawe-Taylor, J. Platt, "Support Vector Method for Novelty Detection," Advances in Neural Information Processing, Vol. 12, pp. 582-588, 2000.

[28] L. Breiman, "Manual for Setting Up, Using, and Understanding Random Forests," V4.0, 2003, URL: www.stat.berkeley.edu/~breiman/Using_random_forests_v4.0.pdf

[29] F.T. Liu, K.M. Ting, Z. Zhou, "Isolation-Based Anomaly Detection," ACM Transactions on Knowledge Discovery from Data, Vol. 6, No. 1, 2012.

[30] I.T. Jolliffe, "Principal Component Analysis," Second Edition, New York: Springer, 2002.

[31] D.E. Denning, "An Intrusion-Detection Model," Proceedings of the IEEE Symposium on Security and Privacy, pp. 118-131, 1986.

[32] H. Javitz, A. Valdes, "The SRI IDES Statistical Anomaly Detector," Proceedings of the IEEE Symposium on Security and Privacy, 1991.

[33] K. Yamanishi, J. Takeuchi, G. Williams, "On-line Unsupervised Outlier Detection Using Finite Mixtures with Discounting Learning Algorithms," Proceedings of SIGKDD, pp. 320-324, 2000.

[34] D. Endler, "Intrusion Detection: Applying Machine Learning to Solaris Audit Data," Proceedings of the 14th Annual Computer Security Applications Conference, IEEE, pp. 268-279, 1998.

[35] R.O. Duda, P.E. Hart, D.G. Stork, "Pattern Classification," 2nd Edition. New York: Wiley, 2000.

[36] LAK, "Loonaangifteketen," 2017, URL: https://www.loonaangifteketen.nl/

[37] R. Foorthuis, "Anomaliedetectie en Patroonherkenning binnen de Loonaangifteketen," Digitale Overheid van de Toekomst, 28 September 2016.

[38] T. Fawcett, "An Introduction to ROC analysis. Pattern Recognition Letters," No. 27, pp. 861-874, 2006.

[39] T. Saito, M. Rehmsmeier, "The Precision-Recall Plot is More Informative than the ROC Plot When Evaluating Binary Classifiers on Imbalanced Datasets," PLOS ONE, March 4 2015.

[40] X. Robin, N. Turck, A. Hainard, N. Tiberti, F. Lisaceck, J. Sanchez, M. Müller, "pROC: An Open-source Package for R and S+ to Analyze and Compare ROC Curves," BMC Bioinformatics, Vol. 12, No. 77, 2011.

[41] S. Macskassy, F. Provost, "Confidence Bands for ROC Curves: Methods and an Empirical Study," Proceedings of the First Workshop on ROC Analysis in AI, ROCAI-2004, 2004.

[42] N. Turck, L. Vutskits, et al., "A Multiparameter Panel Method for Outcome Prediction Following Aneurysmal Subarachnoid Hemorrhage," Intensive Care Med, Vol. 36, pp. 107-115, 2010.

[43] K.H. Zhou, A.J. O'Mally, L. Mauri, "Receiver-Operating Characteristic Analysis for Evaluating Diagnostic Tests and Predictive Models," Circulation, Vol 115, No. 5, pp. 654-657, 2007.

[44] L.L. Pipino, Y.W. Lee, R.Y. Wang, "Data Quality Assessment," Communications of the ACM, Vol. 45, No. 4, pp. 211-218, 2002.

[45] N.G. Weiskopf, C. Weng, "Methods and Dimensions of Electronic Health Record Data Quality Assessment: Enabling Reuse for Clinical Research," Journal of the American Medical Informatics Association, Vol. 20, pp. 144-151, 2013.

[46] M. Onderwater, "Outlier Preservation by Dimensionality Reduction Techniques," International Journal of Data Analysis Techniques and Strategies, Vol. 7, No. 3, pp. 231-252, 2015.

[47] R. Bryll, R. Gutierrez-Osuna, F. Quek, "Attribute Bagging: Improving Accuracy of Classifier Ensembles by Using Random Feature Subsets," Pattern Recognition, Vol. 36, pp. 1291 – 1302, 2003.

[48] T.K. Ho, "The Random Subspace Method for Constructing Decision Forests," IEEE Transactions on Pattern Analysis and Machine Intelligence, Vol. 20, No. 8, pp. 832-844, 1998.

[49] M.L. Shyu, S.C. Chen, K. Sarinnapakorn, L.W. Chang, "A Novel Anomaly Detection Scheme Based on Principal Component Classifier," Proceedings of the ICDM Foundation and New Direction of Data Mining workshop, pp. 172-179, 2003.

[50] P. Helman, J. Bhangoo, "A Statistically Based System for Prioritizing Information Exploration Under Uncertainty," IEEE Transactions on Systems, Man, and Cybernetics, Vol. 27, No. 4, pp. 449 – 466, 1997.

[51] K. Yamanishi, J. Takeuchi, "Discovering Outlier Filtering Rules from Unlabeled Data: Combining a Supervised Learner with an Unsupervised Learner," Proceedings of SIGKDD, pp. 389-394, 2001.

[52] O. Komori, S. Eguchi, "A Boosting Method for Maximizing the Partial Area Under the ROC Curve," BMC Bioinformatics, Vol. 11, No. 314, 2010.

[53] W.J. Youden, "Index for Rating Diagnostic Tests," Cancer, Vol. 3, pp. 32–35, 1950.

[54] H. Liu, F. Hussain, C.L. Tan, M. Dash, "Discretization: An Enabling Technique," Data Mining and Knowledge Discovery, Vol. 6, pp. 393-423, 2002.

[55] R. Foorthuis, "The SECODA Algorithm for the Detection of Anomalies in Sets with Mixed Data," 2017. URL: www.foorthuis.nl

The following page presents the poster for the IEEE DSAA 2017
conference in Tokyo, Japan

# Anomaly Detection with SECODA

Poster presentation of "SECODA: Segmentation- and Combination-Based Detection of Anomalies" at IEEE DSAA 2017

dr. Ralph Foorthuis, UWV, the Netherlands, ralph.foorthuis@uwv.nl

## Introduction

SECODA is a novel general-purpose unsupervised non-parametric anomaly detection (AD) algorithm for datasets containing continuous and categorical attributes. The method is guaranteed to identify cases with unique or sparse combinations of attribute values.

## Typology of anomalies

The typology presents an overview of the types of anomalies. It provides a theoretical and tangible *understanding* of the anomaly types an analyst may encounter. It also aids in *evaluating* which types of anomalies can be detected by a given AD algorithm. The typology differentiates between the set's 'awkward cases' by means of two dimensions: The data types taken into account and the number of attributes analyzed jointly.

|  |  | Nature of the data | |
|---|---|---|---|
|  |  | **Numerical attributes** | **Categorical or mixed attributes** |
| Number of attributes | **Univariate** — Focus on individual attributes (independence) | **Type I** — Extreme value anomaly | **Type II** — Sparse class anomaly |
| | **Multivariate** — Focus on multi-dimensionality (interactions) | **Type III** — Multidimensional numerical anomaly | **Type IV** — Multidimensional mixed data anomaly |

## The SECODA algorithm

SECODA uses the histogram-based approach to assess the density. The concatenation trick – which combines discretized continuous attributes and categorical attributes into a new variable – is used to determine the joint density distribution. In combination with recursive discretization this captures complex relationships between attributes and avoids discretization error. A pruning heuristic as well as exponentially increasing weights and arity are employed to speed up the analysis.

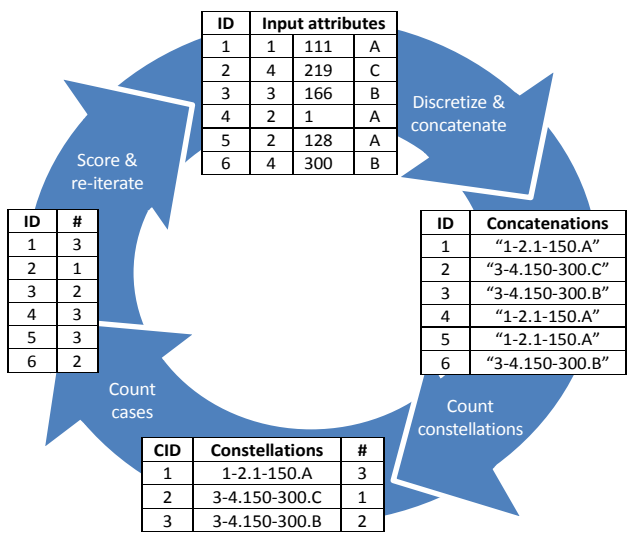

## Evaluation

SECODA was evaluated with multiple simulated and real-world datasets. The diagram below is a 4D snapshot of income data from the Polis Administration, an official national data register in the Netherlands. As can be seen, SECODA was able to detect all four types of anomalies.

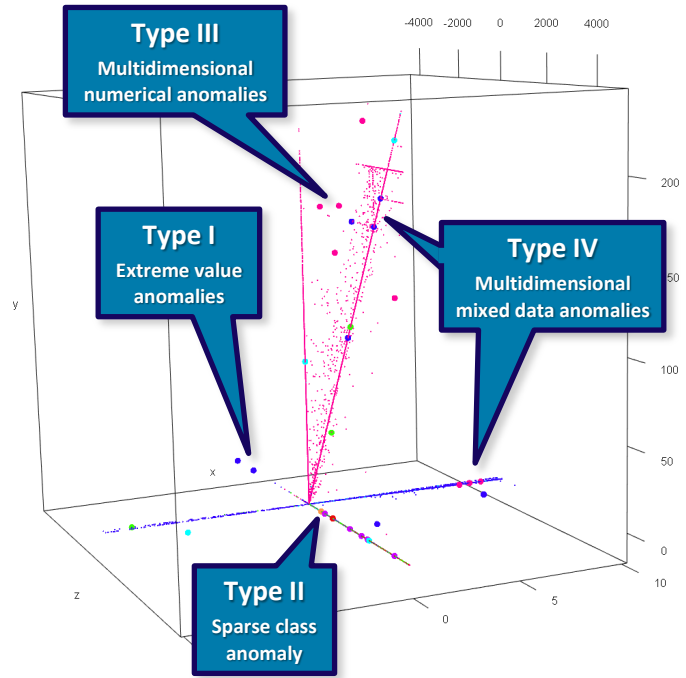

Type III — Multidimensional numerical anomalies
Type I — Extreme value anomalies
Type IV — Multidimensional mixed data anomalies
Type II — Sparse class anomaly

Some characteristics of SECODA:

- Simple algorithm without the need for point-to-point calculations. Only basic data operations are used, making SECODA suitable for sets with large numbers of rows as well as for in-database analytics.
- The pruning heuristic, although simple by design, is a self-regulating mechanism during runtime, dynamically deciding how many cases to discard.
- The exponentially increasing weights both speed up the analysis and prevent bias.
- The algorithm has low memory requirements and scales linearly with dataset size.
- In addition, the real-world data quality use case not only shows that all types of anomalies can be detected, but also that they can be encountered in practice.

Download R code and data examples from www.foorthuis.nl

## Literature

[1] C.C. Aggarwal, "Outlier Analysis," New York: Springer, 2013.
[2] H. Liu, F. Hussain, C.L. Tan, M. Dash, "Discretization: An Enabling Technique," Data Mining and Knowledge Discovery, Vol. 6, 2002.
[3] LAK, "Loonaangifteketen," 2017, URL: www.loonaangifteketen.nl/
[4] R. Foorthuis, "SECODA: Segmentation- and Combination-Based Detection of Anomalies," IEEE DSAA 2017 Conference.